\documentclass[12pt]{article}
\usepackage{graphicx}
\usepackage{amsmath,amssymb}
\usepackage{subfigure}
\usepackage{caption}

\usepackage[a4paper,bindingoffset=0.2in,%
            left=.2 in,right=.2 in,top=1in,bottom=1in,%
            footskip=.25in]{geometry}
\def\bea{\begin{eqnarray}}
\def\eea{\end{eqnarray}}
\def\be{\begin{equation}}
\def\ee{\end{equation}}
\def\nn{\nonumber}

\def\i{\imath}

\def\p{\partial}
\def\tt{\tilde{t}}
\usepackage[utf8]{inputenc} 
\usepackage[T1]{fontenc}
\date{}
\title{Aspects of Quantum Gravity Phenomenology and Astrophysics}
\author{Arundhati Dasgupta and Jos\'{e} Fajardo-Montenegro*  \\ Physics and Astronomy, University of Lethbridge,\\ Lethbridge, Canada T1K 3M4.\\ *Departamento de F\'{\i}sica, Universidad del Valle,\\ Cali 760032, Colombia.}

\begin{document}
\maketitle
\begin{abstract}
With the discovery of gravitational waves, the search for the quantum of gravity, the graviton, is imminent. We discuss the current status of the bounds on graviton mass from experiments as well as the theoretical understanding of these particles. We provide an overview of current experiments in astrophysics such as the search for Hawking radiation in gamma-ray observations and neutrino detectors, which will also shed light on the existence of primordial black holes. Finally, the semiclassical corrections to the image of the event horizon are discussed. \end{abstract}

\section{Introduction}
The gravitational quantum is still elusive experimentally and somewhat ``elusive'' theoretically \cite{grav,mass,pert}. In electrodynamics, the quantum of the electromagnetic wave is known as the photon, and we work with the interactions of photons to derive quantum electrodynamics (QED) phenomena. In the case of gravity, gravitational waves have been discovered 100 years after their prediction. The question is, are there ``gravitons'' or quanta of these waves? Like QED, one can define the ``Fock'' space quantization for the linearized Einstein equations and study free gravitons. However, introducing interactions with gravitons to study scattering amplitudes leads to uncontrollable infinities \cite{pert}. This is known as the ``non-renormalizability'' of perturbative quantum gravity. General relativity might be nonperturbative in the quantum regime, and the story of the quanta could be present in the geometry measurements of area and volume \cite{lqg}. These ``nonperturbative'' theoretical explorations cannot be verified, as they are still in the realm of the microscopic Planck length regime of $10^{-35}$ m. We investigate the semiclassical fluctuations of the flat geometry using loop quantum gravity (LQG) coherent states and discuss whether that can be interpreted as a graviton quantum.

Further in the 1970s, the discovery of black hole thermodynamics and Hawking radiation were studied as ``semiclassical phenomena'', where gravity remained classical and other fields were quantum. The isolated black hole was found to have a temperature proportional to its surface gravity and entropy equal to its horizon surface area. For a solar-mass black hole, which might have formed using stellar collapse, this temperature is of the order of $10^{-8}$ K. If we observe the current-day black holes, then they are immersed in the background cosmic radiation, which has a temperature of  $2.783$ K. As the heat flows from higher to lower temperatures, the black holes would not radiate into the surroundings, and as of now, there is no experimental evidence of Hawking radiation. The study of black hole mergers using gravitational waves has provided evidence for the area increase theorem \cite{area}. How would one obtain a verification of the temperature and radiative properties of black holes? The existence of {\it {primordial black holes}} 
 (PBH) of small mass, originating in density fluctuations of the early universe, would allow for high-temperature black holes and Hawking decays in the form of gamma-ray bursts. The search for PBH has been a subject of experimental study \cite{uni}. We discuss this in some detail, and the approximations which describe the theoretical derivation of Hawking radiation are also discussed. The current experiments provide stringent restrictions on the PBH contributions to photon and neutrino fluxes observed on earth, as well as as fractions of dark matter \cite{pbh01,pbh,pbh0,pbh1,pbh2}. Strangely, new observations from gravitational wave data suggest that there are subsolar mass black holes. Recent work tries to find the origins of these, either as PBH or from other processes without the Chandrasekhar limit in the collapse process \cite{lmbh}. Whereas this is very interesting, this is not exactly the realm of quantum gravity, though the research in this area might shed light on semiclassical aspects.

However, astrophysical phenomena, such as the black hole merger event, the collapse of a supernova to form a black hole, and neutron star mergers, are strong gravitational events. The energies at which the events happen have strongly coupled gravitational interactions. The quantum dynamics near these events is interesting, and even though the effect is weak, one can try and find indirect evidence in the observational data. Using LQG coherent states, some of these can be studied semiclassically. We discuss these and also comment on other observational results from the semiclassical gravity program for astrophysical observations, including that for the image of the event horizon \cite{cohcorr,eht}. There are several collaborations in quantum gravity phenomenology which, in particular, discuss Lorentz violations and quantum anomalies. The appropriate discussions on these topics can be found in \cite{phen}. {For a previous comprehensive review on quantum gravity phenomenology, see \cite{camelia}. One of the aims of this current review is to also provide a pedagogical introduction to some aspects such as the search for primordial black holes, which is a very active \mbox{field currently.} }

This review has discussions on the (i) graviton, (ii) Hawking radiation, and (iii) semiclassical corrections to strong gravity systems such as the event horizon. The following section discusses the theory of the graviton and the experimental bounds. Section \ref{hawk} describes the phenomena of Hawking radiation, as well as the experimental efforts to detect the emitted particles from PBH. Section \ref{event} describes the physics of the event horizon and quantum correction predictions to the same. The final section concludes with the present status of the field of research in the above and future avenues of quantum gravity phenomenology. 

\section{Graviton}
\label{grav}
The electromagnetic (EM) wave is a solution to Maxwell's equation and is observed in nature. The visible spectrum is known as light, the infrared, which we interpret as heat, and radio waves. The ultraviolet radiation is also detectable and useful as are X-rays in many practical day-to-day events. These, when quantized, give us the photon description of the EM wave, and represent the source-free ``free'' EM fields. The actual production of EM radiation is from accelerated charges, but as the waves propagate out in space, they can be studied as ``free'' EM fields. In the case of gravity, Einstein's action is nonlinear, and the gravitational field has self-interactions. To find the ``free'' plane wave which propagates on its own,  we take a linearized gravity, ``weak fluctuations'' over a flat background.  Nonperturbative waves, produced using strong gravitational interactions, have been studied in \cite{miguel}. As the linearized gravitational waves represent classically ``free'' fields, one would expect that the Fock space quantization of these would be obtained similarly to the photon quantum electrodynamics description. However, herein lies the problem: the graviton theory is a nonrenormalizable theory \cite{pert}.  Is it because the graviton vacuum, which represents the Minkowski spacetime is not a vacuum? Is flat space really a vacuum state in a true theory of quantum gravity? Can we have a perturbation over the flat-space system and describe a graviton as a quantum state in the flat-space background? In the case of the EM theory, the EM field propagates in a flat background that, however, serves as a noninteractive arena for the EM fields to propagate. The photon is created and annihilated out of the QED vacuum, which is a state with {the photon quantum number as zero}. In the following, we discuss whether seeking a similar quantum field vacuum for the graviton is relevant. We also discuss the question of which physics of the systems we should experiment for the observation of the graviton.
\subsection{The Linearized Theory of the Graviton}
In the following, we discuss Einstein's theory of the linearized metric. The field equations for the Einstein action is ``free'' in its gauge-fixed form; however, if we try to write the full Einstein Lagrangian for the gravitational field, then there are interaction vertices to all orders for the graviton. The quantum amplitudes including these interactions do not converge, and neither can the theory be renormalized using standard techniques. To begin with, we write the metric of spacetime $g_{\mu \nu}$ as a flat space $\eta_{\mu \nu}$ and a weak fluctuation $h_{\mu \nu}$.
\be
g_{\mu \nu}= \eta_{\mu \nu}+ h_{\mu \nu}.
\ee
It is assumed that $|h_{\mu \nu}|_{\rm max}\ll 1$ ($\mu,\nu,\alpha,\beta\ {\rm etc}=0,\ldots,3$). Note that using standard convention, the metric is dimensionless and the amplitude of the fluctuations are defined using the absolute maximum value.  From experiments \cite{grav}, we are aware now that the amplitude of the ``gravitational wave'' is of the order of $10^{-22}$ as received on earth. One can write the Einstein Lagrangian density as a function of this metric, its determinant $g$, and scalar curvature $R$, 
\be
{\cal L}= \sqrt{g} ~R= -\frac12 \sqrt{-1+h} \left[ (h^{\mu \nu}) (\eta^{\alpha \beta} \partial_{\alpha}\partial_{ \mu} h_{\nu \beta} - \Box\  h_{\mu \nu}) \right]  .
\label{eqn:action}
\ee

In the above, we have kept the terms in the Lagrangian which are quadratic in $h_{\mu \nu}$. The linear terms of the form $\eta^{\mu \nu} \eta^{\lambda \rho}\partial_{\rho}\partial_{\mu} h_{\lambda \nu}$ are total derivatives and contribute only at the boundaries, {which we ignore}. Further,  $\Box\equiv\eta^{\alpha\beta}\partial_{\alpha}\partial_{\beta}$, {and $h$ is the trace of $h_{\mu \nu}$}. The equation of motion from the above to a linear order in ``$h_{\mu \nu}$'' is

\be
\eta^{\alpha \beta} \partial_{\alpha}\partial_{ \mu} h_{\nu \beta} -  \Box\  h_{\mu \nu} =0.
\ee

 This still has a gauge degree of freedom due to diffeomorphism invariance, which can be fixed by putting the $\partial^{\alpha} h_{\alpha \beta}=0$ restriction on the linearized metric.  The equation of motion reduces to a ``wave equation''
 \be
 \Box~ h_{\mu \nu}=0.
 \ee
 
 The solution for this is a transverse wave (due to Lorentz's condition) and has two polarizations as additional restrictions to fix the residual gauge freedom keeping only two \cite{rez}. The two polarizations are taken as $h_+ = A_+ \cos(\omega z -\omega t)$ and $h_{\times}= A_{\times} \cos(\omega z -\omega t)$, if it is propagating in the z-direction \cite{rez}, with angular frequency $\omega$ and amplitude $A_{+}, A_{\times}$.  The question is: can these waves, when quantized, give us ``quanta'' as it is possible for photon quantization? In other words, can one define a Fock space representation for the perturbative Hilbert space of Einstein's gravity?
 The answer is surprisingly difficult, as the Einstein action introduces self-interactions of the gravitons to all orders, which cannot be renormalized using standard field theory techniques. The gravitational propagator can be calculated, but the quantum corrections cannot be made finite using regularization and renormalization techniques. One can see the origin of self-interactions even at this order in the Lagrangian in Equation (\ref{eqn:action}) as the nonpolynomial ``measure'' $\sqrt{-1+h}$ can give rise to the interaction terms upon expanding the square root. A simple ``degree of superficial divergence'' counting of the gravitational perturbative Feynman diagram gives the number as $D=2(k+1)$, where $k$ is the number of independent momentum interactions \cite{dewit}. This number therefore increases with the number of loops in the scattering calculations and cannot be absorbed by redefining the bare Lagrangian. For Yang--Mill's (YM) theory the same degree is given as $D=4-L_e$, where $L_e$ is the number of external legs of the Feynman diagram. The YM theory is therefore renormalizable, as the number of terms in the Lagrangian which need to be renormalized is finite ($0<L_e<4$). One can use asymptotic techniques to obtain a renormalizable effective Lagrangian for gravity, but we do not discuss this in this review \cite{asym}. However, can there be a ``free'' graviton theory where we can ignore all the interactions? Up to a certain length scale, a ``free graviton'' quantization can be formulated, but the entire theory is also complicated by the definition of the ``gravitational vacuum''. In the theory of gravitational physics, the metric is the basic degree of freedom, and the graviton is a ``perturbation'' over the flat-space geometry. In a true quantization of the theory, the flat spacetime geometry is also an emergent ``metric''. If the metric is an operator, then causality and therefore quantization is not defined. The vacuum likely is the state with no metric or the state that is such that
 \be
 \hat{g}_{\mu \nu}\  |0\rangle=0.
 \ee
 
 There have been several attempts to obtain the perturbative quantum state using a polymer state in the nonperturbative quantization framework of loop quantum gravity. We report on those works briefly and then  describe a semiclassical description of a ``gravitational wave'' using LQG. It remains though that the most complicated aspect of Einstein's gravity is the fact that the field which has to be quantized is the metric of the spacetime, the causality of the system is complicated by the quantization, and macroscopic configurations have to \mbox{be emergent.} 
 \subsection{Gravitons in Loop Quantum Gravity}
 It was shown in \cite{ash} that the SU(2) generators of the loop quantum gravity (LQG) variables decouple into three independent gauge generators in the linearized approximation. In LQG, the basic variables are obtained from the ADM formulation of the canonical gravity. The spacetime is foliated by spatial slices $\Sigma$ with a timelike normal vector along the fourth direction, specified using the coordinate $t$. The induced three-metric on $\Sigma_t$ is given as $q_{ab}$, ($a,b=1,2,3$); 
 the metric in the ADM formulation is given as
 \be
 ds^2= - (N^2 +N^aN_a) dt^2 + N^a dx_a dt+ q_{ab} dx^a dx^b,
 \ee
 where $N^2$ is the lapse, $N^a$ is the shift, and $q_{ab}$ is the induced metric of the time slices $\Sigma_t$. The second fundamental form of this metric is $K_{ab}= {\cal L}_t q_{ab}$ and is the extrinsic curvature tensor which characterizes the embedding of the slice.

 The LQG variables are defined using the soldering forms $e^I_a$ which connect the tangent space $(I=1,2,3)$ of the three slices to the world volume. The canonical variables are defined as 
 \be
 e_a^I e_{bI}= q_{ab}, \ \ E^{a}_I E^{b I}= q \ q^{ab}, \ \ \ A_a^I= \Gamma_a^I- K_{ab} E^{b I},
 \label{eqn:grav}
 \ee
 where $e_a^I$ is the triad, $E^a_I$ are densitized triads, and $A_a^I$ have the properties of a connection due to their definition in terms of the spin connection $\Gamma_a^I$ and the extrinsic curvature tensor $K_{ab}$. The details of the variables can be found in \cite{thiem}. There is usually an Immirzi parameter in the definition of the gauge connection, and this reflects an ambiguity in the system. We chose to set it to one, for the purpose of this paper. The internal indices $I$ transform in the SU(2) group, which is isomorphic to the group of rotations in the three-dimensional tangent space \cite{thiem}. The generators of the transformations in the internal directions are the \mbox{Gauss constraints}
 \be
 {\cal G}^I= \partial_a e^{a I} + \epsilon^{IJK} e_{J}^a A_{ a K}.
 \ee
 
 In the linearized approximation, $q=1$, $q^{ab}=\delta^{ab}+h^{ab}$ and $A_a^I=0$, if one keeps the constraint up to a linear order in the fields, the constraint algebra commutes, i.e.,
 \be
 {\cal G}^I_{\rm Lin} = \partial_a (\delta e^{a I}) + \epsilon^{IJK} \delta_J^a \delta A_{a K},
 \ee
 where due to the linearized metric, one has 
 \be
 e^{aI}=\delta^{aI} + \delta e^{aI}, \  \   \   A_{a K}= 0 + \delta A_{a K},
 \ee
 and
 \be
 h^{ab}= \delta e^{a I}\delta^{b}_I, \\
\ee

 \be
 \{\delta e^I_a (x), \delta A_{K b}(y)\}= \kappa \delta^3(x-y) \delta^{I}_{K} \delta_{ab},
\ee
where $\kappa$ is related to Newton's constant $G$ \cite{thiem,hanno}. The $\delta e^I_a$ and the $\delta A_{K b}$ are the linearized dynamical fields, which are quantized. In the limit $\kappa\rightarrow 0$,
\be
 \left\{{\cal G}^I_{\rm Lin},{\cal G}^J_{\rm Lin}\right\}=0.
 \label{eqn:lin1}
 \ee
 
Interestingly, if one keeps the next order in the constraint definition, the algebra is not zero to a linear order as the Poisson bracket gives a linear result in the fields. 
\be
{\cal G}^I_{\rm Lin}= \partial_a (\delta e^{a I}) + \epsilon^{IJK} \left(\delta_J^a + \delta e^{aJ} \right)\delta A_{a K},
 \ee
 and
 \be
 \left\{{\cal G}^I_{\rm Lin},{\cal G}^J_{\rm Lin}\right\}= \kappa\left(\delta A^{IJ}-\delta^{IJ} \delta A^{b}_{b}\right)\delta^3(x-y).
 \ee
 
 This term would go to zero in the limit $\kappa\rightarrow 0$. 
 To avoid these confusions about the algebra and also questions about the Minkowski ``quantum state'' about which perturbation is being performed, we use the full SU(2) degrees of freedom and imposed the linear metric only in the semiclassical approximation. The details of the calculations appear in \cite{cohgrav}. 
 
 For the polymer quantization of linearized gravity using the $U(1)\times U(1)\times U(1)$ Hilbert space, one can use the work of \cite{mad}. This approach is based on the linearized algebra of LQG  variables, as given in Equation \eqref{eqn:lin1}. The LQG phase space thus has a $U(1)\times U(1)\times U(1)$ symmetry in the linearized approximation, instead of the full $SU(2)$. The Hilbert space quantum states are of the form
 \be
 |\vec{\alpha},\{q\}\rangle= |\alpha_1,q_1\rangle|\alpha_2,q_2\rangle|\alpha_3,q_3\rangle,
 \label{eqn:gravstate}
 \ee
 where $|\alpha_i,q_i\rangle$ are elements of a $U(1)$ Hilbert space. $q_{i}$ label integers and $\alpha$ labels the discrete network. The flux operator defined in terms of the triads is given as \cite{mad}
 \be
 X^{a}_{\vec{\alpha},\{q\}(r)}(\vec{x})= \sum_I q_I \int ds_I (\vec{e}_I(s^I),\vec{x}) \dot{e}^a_I,
 \ee
 where $s_I$ is a surface in three dimensions, which the discrete edge $e_I$ of the graph $\alpha$ \mbox{intersects once.}
 
 The Fock space quantum vacuum for the graviton is a transform of the state in \mbox{Equation \eqref{eqn:gravstate}.} Whether this facilitates further study of the perturbation theory of the graviton is yet to be investigated. The transform is given as 
 \be
 \Phi_0 := \sum_{\alpha,{q}} c_{0 \vec{\alpha},\{q\}} \langle\vec{\alpha},{q}|,
 \ee
 where 
 \be
 c_{0 \vec{\alpha}, \{q\}} = \exp\left(- \frac{\imath}{4} \int ~ d^3 x \ G_{ab} ^{\vec{\alpha},\{q\} (r)}(\vec{x}) * X^{ab}_{\vec{\alpha},\{q\}(r)}(\vec{x})\right),
 \ee
 where these are ``smeared'' operators in the LQG polymer space, and $r$ is a measure of the Gaussian smearing ($X_r(\vec{x})= \int d^3 y X(\vec{y}) \exp(-|\vec{x}-\vec{y}|^2/2r^2)/((2 \pi r^2)^{3/2})$). 
 \be
 X^{ab}_{\vec{\alpha},\{q\} (r)}=\sum_i X^a_{\alpha_i,q_i} \delta^b_i.
 \ee
 
 The $G_{ab}^{\vec{\alpha},\{q\} (r)}(\vec{x}) $ is related to the flux of the two ``graviton'' polarizations in the light cone. We refrain from getting into the details of the above, but the reader is urged to follow the details of the derivation in \cite{mad} and \cite{mad1}. Whereas this approach to obtaining a ``quantum'' of linearized gravity is technically rather involved and involves an additional scale ``$r$'' apart from the usual discretization of quantum variables, it is believed to give a polymer representation of the ``graviton''. 
 
 The expectation values of the operators are preserved in the transform and therefore, one loop corrections to the graviton propagator can be tested. A derivation of a one-loop correction using a perturbation of reduced loop quantum cosmology states exists in \cite{gol}. Another reference for the reduced phase-space quantization of linearized gravitational waves is \cite{maj}. Moreover, a more recent work uses the free graviton Lagrangian and obtains a ``polymer state'' for the same. This approach obtains some corrections to the gravitational wave propagator \cite{gw1}. However, in none of the above papers the emergence of the background Minkowski metric is discussed. The self-interaction of gravitons is also not obtained to all orders, as predicted by the Einstein Lagrangian. In the next section, we try to find some phenomenological implications of the graviton's existence in observational data.
 \subsection{Gravitons in Semiclassical Gravity}
 In this subsection, we derive the semiclassical phase space of the gravitational wave metric and obtain a coherent state for the system using the techniques of \cite{thiem,cohgrav}. To begin with, we find the triads for the metric and the LQG gauge connection, which are the classical variables for the system. The details can be found in \cite{cohgrav}. The spatial metric for a standard gravitational wave metric in the tt-gauge is (the lapse is one and shift is zero in the ADM form of the four-metric)
 \be
q_{ab}=\left(\begin{array}{ccc}1+h_+ & h_{\times} & 0 \\ h_{\times} & 1-h_+ & 0 \\ 0& 0 & 1\end{array}\right). \label{metric}
\ee
In the process of obtaining the coherent state for the above metric, we identify the classical phase space in terms of the LQG variables \cite{grq}.
The triads $e_a^I e_{b I}=q_{ab}$ are obtained as

{\small \be
{e}{^{I}_a}=\left(\begin{array}{ccc}\sqrt{\frac{1-(h_+^2+h_{\times}^2)}{2(1-h_{\times})}} & \sqrt{\frac{1-(h_+^2+h_{\times}^2)}{2(1-h_{\times})}} & 0 \\ \frac{1-(h_{\times}-h_+)}{\sqrt{2(1-h_{\times}})} & \frac{-1+(h_{\times}+h_+)}{\sqrt{2(1-h_{\times}})} & 0 \\ 0 &0 &1\end{array}\right) =\left(\begin{array}{ccc}\frac{1}{\sqrt{2}}+\frac{h_{\times}}{2\sqrt{2}} & \frac{1}{\sqrt{2}}+\frac{h_{\times}}{2\sqrt{2}} & 0 \\
              \frac{1}{\sqrt{2}}+\frac{1}{\sqrt{2}}(h_+-\frac{h_{\times}}{2})& -\frac{1}{\sqrt{2}}+\frac{1}{\sqrt{2}}(h_++\frac{h_{\times}}{2}) & 0 \\
              0 &0 & 1\end{array}\right).
\ee}

Obviously, in our gauge choice, the triad is not diagonal at the zeroth order.
The extrinsic curvature of the metric is obtained using the definition $K_{ab}= -\partial_t q_{ab}$, and the SU(2)-valued gauge connections defined in Equation (\ref{eqn:grav}) are:
\bea
{A}{^{{1}}_x}&=&-\frac{1}{2\sqrt{2}}(\partial_z h_{\times}+\partial_z h_+)={A}{^{{2}}_y} \nn \\
{A}{^{{1}}_y}&=&-\frac{1}{2\sqrt{2}}(\partial_z h_{\times}-\partial_z h_+)=-{A}{^{2}_x} \nn \\
{A}{^{{1}}_z}&=&{A}{^{{2}}_z}={A}{^{3}_x}={A}{^{{3}}_y}=0 \nn \\
{A}{^{{3}}_z}&=&\frac{1}{2}\partial_z h_+. \nn
\eea

We also computed the nonzero spin connections for this metric \cite{grq}. 
 Next, we take a discretization of the background geometry. This smearing of variables is required to obtain smooth commutators of the quantum theory, instead of distributional delta functions. For details, see \cite{thiem}, and the smearing of the gauge connection on one-dimensional curves gives holonomies which involve path-ordering. 
 \be
 h_e(A)= {\cal P} \exp\left(\int A\right).
 \ee
 
\begin{figure}
     \centering
     \includegraphics[scale=0.95]{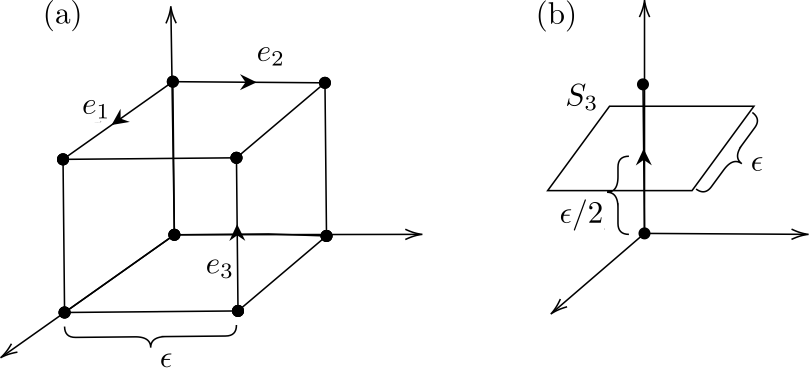}
     \caption{(\textbf{a}) Building block for the decomposition of the 3-geometry. (\textbf{b}) Example of one of the smearing surfaces to calculate the momenta.}
     \label{fig:cubic}
 \end{figure}

 The discretization is not dictated by the theory but is motivated from the flat geometry of the classical three-metric. We take a planar graph, which form a cubic 3-d polyhedronal decomposition of the three-geometry, as shown in Figure \ref{fig:cubic}. Therefore, there are six links and/or six faces meeting at a given vertex.
 The holonomies and the momentum are calculated as smeared along the one-dimensional edges of the graph, and the two-dimensional faces of the cube which the links intersect precisely at one point. These calculations are done using the techniques of \cite{cohgrav}. The holonomies of the three independent links in the $x$, $y$, and $z$ directions and the corresponding momenta are given up to a linear order 
 in the amplitudes $A_+$, $A_{\times}$,
 \bea
 h_{e_x} &= & 1-i \frac{\epsilon}{2} A_x^I \sigma_I\\
 h_{e_y} & = & 1 - i \frac{\epsilon}{2} A^I_y\sigma^I\\
 h_{e_z} & =& 1+i \frac{A_+}{2} \sin\left(\omega\left(z_0-t_0 + \frac{\epsilon}{2}\right)\right)\sin\left(\frac{\epsilon}{2}\right) \sigma_3,
 \eea
where one has taken a vertex at $(x_0,y_0,z_0)$ and the links are of width $\epsilon$. $\sigma_I$ are the Pauli matrices. Next, one takes the faces centred at the middle of the links, i.e., at $x_0+\epsilon/2$, $y_0+\epsilon/2$, and $z_0+\epsilon/2$, and of area $\epsilon^2$. The momenta are labelled by the edges which intersect the faces. The momenta are defined as 
 $P^I_e=\frac1{\kappa}\int_{S_e} * E^I $.
 \bea
 P_{e_x}^1&=&\frac{1}{\sqrt{2} \kappa }\left(\epsilon^2 +\frac{\epsilon^2 ( A_{\times})}{2}\cos(\omega(z_0-t_0))\right)\\
 P_{e_x}^2&=& \frac{1}{\sqrt{2} \kappa }\left(\epsilon^2 +\frac{\epsilon^2 (2 A_+- A_{\times})}{2}\cos(\omega(z_0-t_0))\right)\\
 P_{e_y}^2& =&\frac{1}{\sqrt{2} \kappa }\left(-\epsilon^2 +\frac{\epsilon^2 (2 A_++ A_{\times})}{2}\cos(\omega(z_0-t_0))\right)\\
 P_{e_y}^1&=&\frac{1}{\sqrt{2} \kappa }\left(\epsilon^2 +\frac{\epsilon^2 ( A_{\times})}{2}\cos(\omega(z_0-t_0))\right)\\
 P_{e_z}^3 &= &\frac{1}{\kappa}\epsilon^2.
 \eea
 
 As the densitized triads are smeared over two-dimensional areas and acquire dimensions, the momenta are defined with the dimensional constant $1/\kappa$, $\kappa= 8 \pi G/c^3$ to make the variables dimensionless. In the quantum version, this acquires the role of $1/\hbar \kappa= 1/l_p^2$, where $l_p$ is the Planck length.
 The coherent states are defined as peaked at the classical values of a complexified SL(2,C) element as specified by Hall \cite{hall}, $$g_e= \exp(i T^I P_e^I) h_e,$$ and a detailed coherent state can be written for the above classical phase space, now described only using the discrete one-dimensional smeared holonomies and corresponding momenta. Note these ``coherent states'', as defined in \cite{thiem} for LQG, are representative semiclassical states and are not exactly identifiable as ``coherent states'' as in completely solvable Hamiltonian systems. However, these states have minimal uncertainty in the time slice they are defined in. Next, we calculate the semiclassical corrections to the geometry by using the results of \cite{cohcorr}. The coherent states are given for one such discrete element $e$ and the LQG smeared variables as,
 \be
 \psi^t(g_e,h_e)=\sum_j (2j+1) \exp(-\tilde{t} j(j+1)/2) \chi_j(g_e h_e^{-1}),
 \ee
 where $\chi_j(h_e)$ is the character of the $j$th irreducible representation of SU(2). One can find the expectation value of the momentum operator $\hat{P}_e^I$ in this state, and one obtains it to the first order 
 in the semiclassical parameter $\tilde{t}$ \cite{cohcorr}
 \be
 \langle\psi^t|\hat{P}_e^I|\psi^t\rangle= P^I_e\left(1+\frac{\tilde{t}}{P_e}\left(\frac1{P_e}-\coth(P_e)\right)\right)=P^I_e\left(1+\tilde{t} f(P_e)\right),
 \label{eqn:corr}
 \ee
 where $P_e= \sqrt{P_e^I P_e^I}$ and $f(p)=(1/p)(1/p-\coth(p))$. Therefore, one can calculate the semiclassical corrections to the metric of the classical gravitational wave, if one writes a coherent state for each discrete element $e$ which comprises the entire Minkowski three-volume divided into cubic cells as in the figure. The vertices of the cube which are shared by three+three coherent states and these can have SU(2) intertwiners \cite{sigma}, but the nature of the corrections remain the same. Note these coherent states are not exactly similar to the coherent states for photons, which are Abelian. These coherent states are non-Abelian \mbox{in nature.}  
 
 In fact, if we take the pure Minkowski space and use the coherent state as a measure of the quantum fluctuation, what would we generate as the corrected metric? All the $P^I_e$'s for the Minkowski metric can be obtained as given above and, in the limit, $A_{+,\times}=0$ would represent the Minkowski metric. In this particular gauge, the corrections generate semiclassical fluctuations in the $\eta_{xx}$, $\eta_{yy}$, and $\eta_{zz}$ components but not in the $\eta_{xy}$ directions. 
 
 Next, we discuss the fluctuations to the gravitational wave metric as generated from the coherent state which peaks at the gravitational wave metric. Obviously, the metric would fluctuate and generate semiclassical corrections to the geometry at order $\tilde{t}$. We set the semiclassical parameter (which has to be dimensionless) as a ratio of the Planck scale to the gravitational wave, wavelength, or $\tilde t= l_p^2/\lambda^2$. We take the wavelength as that is the length scale which characterizes the wave system. A relevant-frequency gravitational wave, which might generate detectable semiclassical fluctuations, has to be of very high frequency. Let us say a $10^{35}$ Hz gravitational wave will have the semiclassical parameter as $\tilde{t}\approx 10^{-16}$.
 
 In the above, have we predicted a ``quantum origin'' of the gravitational wave that would comprise the ``graviton''? Obviously, the story is not about particles in gravitational physics, or matter quanta, but the quantum of geometry. The tiny area measurements in each basis state of the operator $\hat{P}^I_e$ represent the ``graviton'', the condensate of which is represented by the coherent-state wave packet. It thus remains that from our perspective, the Minkowski geometry is not the gravitational vacuum, but also emergent from a semiclassical state. Therefore, one should not confuse the quantum gravity vacuum state with the ``matter vacua''. 
 
 We suggest two ways to search for quantum gravity bounds/origins in a gravitational wave experiment:
 \begin{itemize}
 \item [(i)] As the coherent states are non-Abelian in nature, the expectation values of operators have semiclassical corrections which originate due to self-interactions. These can be detected for high-frequency gravitational waves. 
 \item [(ii)] The search for individual ``gravitons'' or quanta of geometry would require much more precise instruments, able to resolve the coarse-graining of geometry itself.
\end{itemize} 

  The latter (ii) will require further investigations, in particular about what the dynamical fundamental ``quanta'' of LQG is.  One also has to find if there is a gauge invariant observable which is measurable in experiments. Our questions seem to seek answers by quantizing matter and the gravitational degrees of freedom simultaneously. {\it {However, due to the hierarchy problem, it is preferred that matter is quantized and the gravitational degrees of freedom are semiclassical in the current epoch.}}
    In the combined Hilbert space of the matter and gravitational degrees of freedom $H_{\rm matter}\otimes H_{\rm grav}$, the combined matter--gravity state should be taken as 
 \be
 |\Psi\rangle= |\psi_{\rm matter}\rangle \otimes |\psi^{\rm grav}_{\rm semiclassical}\rangle.
 \ee
 
 For previous work in adding matter interactions in LQG, refer to \cite{hanno}.

 Using criterion (i) and the idea that matter quanta interact with gravitational degrees of freedom at semiclassical length scales, one finds that the semiclassical fluctuations of the metric are relevant. We therefore calculate the metric corrections as predicted from the coherent states for LQG constructed by Thiemann, Winkler, \cite{thiem} and as observed in \cite{cohcorr}. They emerge as
 \bea
 g_{xx} &= &( 1 + h_+)(1 + 2\tilde{t} \ f(P_{e_x}))\\
 g_{yy} & =& (1-h_+)(1 + 2 \tilde{t} \ f(P_{e_y}))\\
 g_{xy} &=& h_{\times}(1 + \tilde{t} \  f(P_{e_x}) + \tilde{t} \ f(P_{e_y}))\\
 g_{zz} &=& 1 + \tilde{t}\ f(P_{e_z}).
 \eea
 
The gauge invariant momenta are found to be:
\bea
P_{e_x} &=& \frac{\epsilon^2}{\kappa}\left(1+ \frac12 h_+\right)\\
P_{e_y} &=& \frac{\epsilon^2}{\kappa}\left(1-\frac12 h_+\right)\\
P_{e_z} &=& \frac{\epsilon^2}{\kappa}.
\eea

The continuum limit is obtained using $\lim_{\epsilon\to 0} P_e/\epsilon^2$. This gives the metric fluctuations at a location $(x_0,y_0,z_0)$ and one can solve the propagation of matter in this corrected metric. As evident in the continuum limit, the corrections are functions of the classical triads, and thus dependent only on the $z$ coordinate. Moreover, the corrections are relevant only at one instant $t=t_0$ of the spacetime. For a 100 Hz frequency, the gravitational wave will have a semiclassical correction of the order of $10^{-84}$, which is way smaller than the gravitational wave amplitude. If one probes higher-frequency gravitational waves, and therefore shorter wavelengths, the Planck scale coarse-graining will start manifesting itself and the effects might be evident in a gravitational wave detector. The Minkowski metric is also corrected semiclassically, and one can probe these using quantum fields in these geometries.
\subsection{Summary} In this section, we gave a ``semiclassical'' state which could describe a gravitational wave at one instant. It predicted fluctuations which could be measurable for high-frequency waves $\geq 10^{30}$ Hz. These frequencies were way above the ones observed in the LIGO detectors. From the current observation of gravitational waves, there are bounds on the ``graviton mass''. From LIGO, the bound is $1.2 \times 10^{-22} \ {\rm eV}$. This bound does not shed light on the origins of the mass from the methodology. Theoretically, the graviton mass can originate from quantum corrections to the Einstein theory, as well as from matter interactions which preserve diffeomorphism invariance. In this review, we do not discuss massive gravitons. 
\section{Search for Hawking Radiation and Primordial Black Holes}
\label{hawk}
The discovery that quantum mechanics near black hole horizons results in particle creation originates in the paper by SW Hawking \cite{cmp}. In that paper, a quantum field vacuum was time-evolved in the collapsing geometry of a star. The quantum state evolved into a thermal state, with a temperature inversely proportional to the mass of the black hole. In \cite{cmp}, it was shown that the exact temperature of a solar-mass black hole was $10^{-8}$ K. However, it would not radiate into the surrounding, which was at $2.78$ K. This led to the search for black holes with mass $\sim 10^{14} \, {\rm g}$, and these could have formed in the early universe. Due to the Chandrasekhar limit, astrophysical black holes have a bounded mass if formed from stellar collapse.
On the other hand, early universe density fluctuations can lead to the formation of tiny black holes, with horizon size fractions of a millimetre. These black holes have intrinsic temperatures higher than the current CMB temperature of $2.78$ K. Even if the early universe had been hot, as the primordial universe cooled down, these black holes would start radiating and evaporate eventually or form Planck size remnants. 

\subsection{Formation of Primordial Black Holes (PBH)}
The story of the collapse of matter to form black holes is well-studied in the work of Choptuik \cite{bhf}. Scalar data in an initial slice undergo collapse, and the mass of the black hole formed has a scaling equation. This physics is true for early universe cosmology. It is noted that the matter undergoing collapse is taken as dust in most calculations and the Fermion/quark composition (required for the Chandrasekhar limit) of the cosmic soup is mostly ignored. For a comprehensive review of primordial black hole formation, one is referred to \cite{uni}. Here, we briefly outline the methods used to study matter collapse in the early universe. One of the main ingredient in the study of collapse in the early universe is Jean's instability. This instability characterizes density fluctuations in a fluid. The formula for Jean's instability is obtained by equating the time for free fall (or the time taken for an object of radius $R$ to collapse under its own gravity) to the time taken by a sound wave to cross the radius. It is therefore a critical radius for which a pressure wave in the fluid gets trapped. Jean's critical length can also be obtained by solving for perturbations flowing in a fluid and the self-gravitational force generated by the perturbation. In the following, we discuss Jean's instability.

\subsection{Jean's Instability}
In this section, we discuss the collapse in a fluid of density $\rho$. This process also gives a rough description of the physics of a ``density'' collapsing under ``perturbations'' or under its own weight.
The time for ``free fall'' of a mass in an elliptic orbit of eccentricity one, according to Kepler's laws (of planetary motion) is
\be
\tau^2= \frac{\pi^2}{2}\frac{R^3}{{GM}}, 
\ee
where $M$ is the mass causing the orbit, and $R$ is the distance from the focus of the ellipse. We use this to model self-collapse of a mass under its own gravity. If the mass collapses, then only half of this time is taken. Given that the total mass in a radius $R$ of a spherical distribution of constant density $\rho$ is 
\be
M= \frac{4\pi}{3} R^3 \rho,
\ee
approximating the mass using this formula, the time for free fall is given as a function of density as 
\be
\tau= \sqrt{\frac{3 \pi}{32 G \rho}}.
\ee

If the speed of sound in the fluid is $c_s$, then the time for sound to flow through a distance $R$ is 
\be
\frac{R}{c_s}.
\ee

This time would be the same as that a pressure wave flowing through the medium would take. If the gravitational collapse time is greater than the pressure wave time, the mass is unstable, and the critical length scale of the fluid region is given as
\be 
R_{\rm JL}= \left(\frac{3\pi}{32}\right)^{1/2} \frac{c_s}{\sqrt{G\rho}}.
\ee

The same ``collapse formula'' can be derived using a spherical homogeneous mass $M$, whose radius increases by a perturbation $\Delta R=-\alpha R$, where $\alpha$ is a small perturbation.  The change in pressure using the formula $\delta p/\delta \rho=c_s^2$ can be related to the change in density due to the compression, and this gives rise to a force and ``acceleration'' \mbox{obtained as}
\be
a_p= \frac{\delta p}{\rho_0 R}=\frac{3\alpha c_s^2}{R}.
\ee

In the above, we took $\delta\rho= 3 \alpha \rho_0$, where $\rho_0$ is the original density. Simultaneously the shrinking of the radius gives rise to an increase of the Newtonian acceleration
\be
a_{\rm g}= \frac{2 GM \alpha}{R^2}.
\ee

If the gravitational acceleration exceeds the ``pressure acceleration'', the mass is expected to collapse, which gives a critical length
\be
\frac{3\alpha c_s^2}{R_C}= \frac{2 GM \alpha}{R_C^2}=\frac{4\pi}{3}\rho_0R_C^3\frac{2G}{R_C^2}\rightarrow R_c\propto \frac{c_s}{\sqrt{\rho_0 G}}.
\ee

Thus, the critical radius for the collapse in a fluid of density $\rho$ is proportional to the speed of pressure waves $c_s$ in the medium. Here, one of the important assumptions for the calculation of the speed of sound is the assumption that for the early universe fluid, entropy is conserved. We next discuss if a change in the description of the fluid of the early universe might change this Jean's length. The above discussion on Jean's instability can be found in many references, including \cite{jeans2,jeans1}.

\subsection{A Quantum Entropy Production Fluid and Jean's Instability} 
In the above Newtonian derivation of gravitational collapse, the requirement that the fluid be isentropic may not be true in the early universe. In fact, entropy production causes the flow of the universe to be as in an ``open system'', where the big bang singularity is resolved \cite{cosmop}. We take a slight detour and discuss the situation where there is entropy production in the fluid as anticipated in \cite{cosmop}. In \cite{cosmop}, it is conjectured that spacetime can generate particles which add to the fluid, the energy momentum tensor of the Einstein equation. This particle creation is a quantum process and might add insight to the origins of today's cosmological observations. {In \cite{cosmop}, it is shown that in such open systems, cosmological singularity is not formed. In this review, we briefly discuss whether the open system allows for PBH formation.} The conservation law for open thermodynamic systems is given as 
\be
d(\rho V) + p dV - \frac{h}{n} d(n V)=0,
\ee
where $n$ is the particle number and $h= \rho+p$ is the ``enthalpy'' of the system. In most irreversible systems, as in systems with chemical reactions, enthalpy is a measure of the energy of the system, and is a path-independent quantity. The thermodynamics of these systems is controlled by the chemical potential $\mu$, and the entropy per unit volume ``$s$'' is defined as
\be
\mu n = h-Ts,
\ee
with $T$ being the temperature of the system. 

The pressure for this fluid is given as
\be
p=\frac{n \dot{\rho}}{\dot{n}}-\dot{\rho}.
\label{eqn:pressure}
\ee

If one assumes a fluid in the form of ``radiation'', i.e., $\rho = aT^4$ and $n= b T^3$, where $a$ and $b$ are dimensional constants \cite{cosmop}, obviously, from Equation (\ref{eqn:pressure}), the equation of state is $p=\rho/3$. In such an open system, if one obtains the propagation equation of a ``pressure wave'', then the conservation of mass and momentum equations are different. In previous work, the speed of sound in such a fluid was taken as $c_s=\sqrt{1/3}$, which was at constant entropy for the calculation of the Jean's instability. However, the speed of sound changes in a fluid with entropy production. We try to see the origin of the speed of a pressure wave in a gravitating fluid, and it is nonisentropic, with dynamics given by the equations above.
To describe the propagation of pressure waves in a system, one uses the following equations:
For the conservation of mass equation in the fluid, one has
\be
\frac{\partial \rho}{\partial t} + \vec{\nabla} \cdot(\rho \vec{v})= \dot{n_i},
\ee
where we have the ``convective'' derivative of the density and any particle production on the other side of the equation. The conservation of momentum equation or Euler's equation gives (we assume that the fluid is not viscous)
\be
\frac{\partial (\rho \vec{v})}{\partial t} + \vec{v} \cdot \vec{\nabla} (\rho \vec{v}) = - \vec{\nabla} {p} +\rho g.
\ee

In the above, the Navier--Stokes equations have been reduced by setting the viscosity to zero. On the right-hand side, there is a potential term which can be a gravitational potential term.  In all discussions for the speed of sound, or the speed of pressure waves in the system, the velocity is taken to be small, and the density and pressure undergo perturbations. 
We assume no gravitational potential at this stage. If there is a linear perturbation in the velocity, density, and pressure of the fluid, with the $\dot{n}$ remaining the same, the perturbations lead to the following equations
\be
\frac{\partial \delta \rho}{\partial t} + \rho_0 \vec\nabla \cdot \vec{\delta v}=0,
\label{eqn:linperp}
\ee
and
\be
\rho_0 \frac{\partial \vec{\delta v}}{\partial t} = -\vec{\nabla}\delta p.
\label{eqn:linperp2}
\ee

If the system is isentropic, i.e., homogeneous, one can take a partial derivative of Equation (\ref{eqn:linperp}) and obtain
\be
\frac{\partial^2 \delta \rho}{\partial t^2} + \rho_0 \vec\nabla \cdot \frac{\partial \vec{\delta v}}{\partial t}=0.
\ee

In the above, using Equation (\ref{eqn:linperp2}), one obtains
\be 
\frac{\partial^2 \delta \rho}{\partial t^2}- \nabla^2 \delta p=0.
\ee

In the isentropic approximation 
\be
\delta \rho= \left(\frac{\partial \rho_0}{\p p_0}\right)_s \delta p,
\ee
one plugs in the above and obtain
\be 
\frac{\partial^2 \delta \rho}{\partial t^2}- c_s^2 \nabla^2 \delta \rho=0,
\ee
and one obtains the speed of propagation of the density perturbations as
\be
\frac{1}{c_s}= \sqrt{\left(\frac{\partial \rho_0}{\p p_0}\right)_s }.
\ee

In case the fluid has entropy changes, they induce a change in volume. One therefore can obtain for nonisentropic fluids
\be
\delta \rho= \left(\frac{\partial \rho_0}{\p p_0}\right)_s \delta p+ \left(\frac{\partial \rho_0}{\p s_0}\right)_p \delta s.
\ee

If we use the thermodynamic equation for entropy production as
\be
\delta s =  \left(\frac{\partial s_0}{\p \rho_0}\right)_T\delta p,
\ee
then, in the formula for the ``density perturbation'' velocity, we have
\be
c= \sqrt{\frac{c^2_sc^2_p}{c_s^2+c_p^2}},
\label{eqn:reffluid}
\ee
where 
\be
\frac1{c_p^2} = \left(\frac{\partial \rho_0}{\p s_0}\right)_p\left(\frac{\partial s_0}{\p p_0}\right)_T.
\ee
 
 If we add the gravitational potential in Euler's equation, then the wave equation has an inhomogeneous term which has a ``force driving term'' obtained from the gradient of a gravitational potential. If we take the potential to originate from the density, we have $\nabla^2 \phi_1= 4 \pi G \rho_0$, then
 \be 
\frac{\partial^2 \delta \rho}{\partial t^2}- c^2 \nabla^2 \delta \rho=- 4\pi G \rho_0 \delta\rho.
\ee

 We assume a plane wave solution for the density wave $\delta\rho\sim e^{i (\omega t + \vec{k}\cdot\vec{x})}$, and we find
\be
\omega^2 - c^2 k^2 = 4 \pi G \rho_0,
\ee
so a critical ``pressure wave'' is identified. For waves with wave numbers above that, the system will see instability. The critical wave number is given as
\be
k^2=\frac{4 \pi G \rho_0}{c^2}.
\ee

Jean's instability is thus identified as perturbations having a wavelength greater than 
\be
\lambda_J> \sqrt{\frac{\pi}{G\rho_0}} \, c.
\label{eqn:inst1}
\ee

Unlike the previous estimate of the length scale where the gravitational instability sets in, here, the speed of sound is not a mere $\sqrt{1/3}$ as given in the formula for an isentropic radiation fluid but is obtained using Equation (\ref{eqn:reffluid}). In a turbulent early universe, therefore, it is expected that the fluid would be nonisentropic. In addition, the open universe will ensure entropy production as spacetime generates particle species to add to the fluid. As the speed differs, so will the threshold for the formation of PBH. Note the origin of this change from an underlying quantum theory is implicit in the velocity change of the pressure wave. Note our results for a nonisentropic fluid is just one way to see how some of the formulas used for PBH might change; for other origins of change in Jean's instability formula in cosmic fluids, see \cite{jsc}.
\subsection{PBH Formation}
How does one obtain the dynamics of formation of PBH in the early universe? It is postulated that the FLRW universe metric could have perturbations induced by the density fluctuations of the fluid. These can be modelled using a spherical symmetry, and the conditions for the formation of ``trapped surfaces'' or apparent horizons derived using the ``Misner--Sharp'' equations. These PBH can then accrete and grow in size, and there can be PBH formed of masses which are bigger than the solar masses of $10 M_{\circ}$--$30 M_{\circ}$. 
A great deal of the current work on PBH discusses these and the fraction of PBH contributing to dark matter halos $f_{PBH}$. For further reading on the PBH production and the interest in them as contributors to dark matter and physical processes such as microlensing, etc., refer to \cite{pbh}.  
As the black hole formation follows the same numerical flow as in the spherical collapse obtained by Choptuik, the PBH's mass has the following ``scaling'' formula
\be
M_{\rm PBH} = K\  M_{H} (t_H) (\delta_m-\delta_c)^{\gamma},
\ee
where $\delta_m= (\rho-\rho_b)/\rho_b$ is the fluctuation in the fluid density over the Hubble density, at the radius where a compaction function is maximum. $\delta_c$ is the fluctuation at the critical radius related to the Jean's instability in the fluid found earlier. $\delta_c$ represents the threshold of black hole formation. This equation can only be trusted in the regime $\delta_m-\delta_c\sim 10^{-2}$. $M_H(t_H)$ is the Misner--Sharp mass of the horizon, $K$ is a numerical constant. $\gamma$ is a universal scaling exponent and varies depending on the fluctuation profile and the equation of state of the fluid. This equation provides the basis for PBH formation, though using classical equations. The compaction function $C(r,t)$ is defined as the excess of mass over the FLRW mass $M_b$ defined as $M_b= 4 \pi \rho_b R^3/3$,
\be
C(r,t)= 2 \frac{M(r,t)-M_b(r,t)}{R(r,t)}.
\ee

If one takes the perturbation of the FLRW metric to be modelled by a function $\zeta(r,t)$, in the FLRW metric three-slice as $a^2(t) e^{2 \zeta(r,t)} r^2 d\Omega$, one gets a formula for the compaction function in terms of this parameterized fluctuation as
\be
C(r)=\frac23\left(1-(1-r \zeta'(r))^2\right).
\ee

This facilitates the study of this function in terms of the curvature fluctuations of the metric. The various calculations of the ``peak'' values of this compaction function use different ensembles for the fluctuations and accordingly, obtain different values. It is postulated that when the compaction function exceeds a critical value, a collapse occurs, otherwise the fluctuation dissipates away.  The density contrast parameter is related to the peak value of the compaction function as 
\be
\delta_m= C(r_m).
\ee

In this article, we refrain from discussing the various ways of finding PBH compaction function but only show a way the change in threshold value $\delta_c$ of PBH formation influences the collapse process. This critical value is related to Jean's instability in the cosmic fluid and as shown previously, vary according to the approximations used.  A dependence on the formula for PBH on the nature of the fluid is discussed in \cite{pbh}. 
As shown in Equations (\ref{eqn:inst1}) and (\ref{eqn:reffluid}), the threshold of the onset of the instability of a fluid changes if quantum ``particle creation'' is allowed. In \cite{cosmop}, the fluid exchanges particles with the gravitational ``quantum field''. In this open universe, there is no initial singularity \cite{cosmop}, and as we anticipate, the formation of PBH would also differ. The masses would be different, and the nature of the cosmological fluctuations of the gravitational metric would also differ as per the ``entropy production'' of this open universe. A more detailed calculation using quantum cosmology is required for the exact changes required in the {\it theoretical} predictions of the PBH's mass, and the PBH formation from the cosmic soup.

 The formation of PBH can vary from masses of the order of $10^{5}$ g - $ 10^{50}$ g, and therefore, they can range from small black holes to larger-than-solar-mass black holes. The lower limit is based on the Planck mass and the upper limit is based on the cosmological mass. How can we verify the existence of PBH? The existence of PBH can be verified using the observation of particles received on earth, which might have originated from the PBH using the Hawking radiation process. It is this process which we describe next. We discuss PBH whose evaporation time $\propto M^3$ is about the age of the universe. These PBH might have radiated away their mass in the form of photons and neutrinos and would provide evidence for the phenomena of Hawking radiation. The mass of these black holes is estimated as $< 10^{14} {\rm g}$.

Curiously, there was an attempt to find quantum gravity effects on PBH production using loop quantum cosmology (LQC) corrections to the scale factor and the density \cite{dwi}. The authors found that using the LQC-corrected early universe cosmology, the production of PBH was increased theoretically compared to estimates from other theoretical models as that of the Brans--Dicke gravity.
\subsection{Evaporation of PBH}

The mechanism of radiation from black holes can be studied using the power law for the emission of particles. In the 1970s \cite{cmp,page,astro}, one typically calculated the power law using Hawking's formula for the particle flux from black holes. The total energy radiated per unit time from PBH of Hawking temperature $T_H$ is given as
\be
\frac{dE}{dt}= \int d\omega \int d\Omega \sum_{lm} \frac{\Gamma_{\omega slm}}{\exp(\omega/T_H)\pm 1}
\label{eqn:hawk}
\ee
where $\Gamma_{\omega slm}$ is the grey-body factor for the black hole geometry and represents matter waves scattering off the gravitational potential outside the black hole. $s,l,m$ represent the spin and angular momentum quantum numbers of particles with frequency $\omega$.  The sign in the denominator is positive for bosons and negative for fermions. The Hawking temperature for a nonrotating black hole is inversely proportional to the mass. The grey-body factor is calculated using the solutions to the classical equation of motion of the particles in the black hole background and is a function of the spin, angular momentum, mass, and frequency of the emission. The fraction of power radiated in different species can be calculated. The total power radiated can be calculated numerically as 
\be
P= 2.011 \times 10^{-4} \, \hbar c^5 G^{-2} M^{-2},
\ee
where $M$ is the mass of the black hole. Most of the above is radiated out in the form of neutrinos (81.4\%), 16.7\% as photons and 1.9\% as gravitons, as long as the black holes have mass $M>10^{17} {\rm g}$ \cite{page} 
After the black hole has shrunk further, the temperature being higher, and the mass being denser, the black hole radiates quarks in the form of muons and other particles such as electrons and positrons. For this range of black holes, $10^{14}$ g $< M < 10^{17}$ g the power radiated was found to be 
\be
P= 3.6 \times 10^{-4} \, \hbar c^5 G^{-2} M^{-2},
\ee
 90$\%$ is equally divided in electrons, positrons, and neutrinos, 9$\%$ in photons, and 1$\%$ in gravitons \cite{page}. In this work, when computing the power of Hawking particles, the numerical calculations of the grey-body factors were used, and the above division into fractions were based on the spin of the particles. The emission of massive particles would have a different calculation, but for a detection on earth, the massless particles acquire relevance.

In a follow up work \cite{astro}, the emission of gamma rays with energy of about 120 MeV was discussed, and a study of ``gamma ray bursts'' from evaporating PBH was introduced. In there, a mass distribution was assumed for PBH, and this is an ingredient in the current analysis of the data received on earth.
The search for Hawking radiation phenomena in the universe is thus a search for primordial black holes and the particles emitted from them. There are several searches for primordial black holes using gamma-ray bursts which might be the evidence of these black holes evaporating. In the next, we describe some of these searches in detail and provide a bibliography.

\subsection{Archived Data} The Imaging Compton Telescope (COMPTEL) \cite{comp} was decommissioned in 2007, but there remained the archived data to analyze gamma rays. {The search from these data has shown bounds for the primordial black holes (PBH) <$10^{17} {\rm g}$ \cite{comp2}}.

\subsection{Gamma-Ray Bursts} There are several satellite-based experiments, which are functional or at the planning stage such as AMEGO and e-ASTROGRAM. AMEGO is an abbreviation for the All-sky Medium Energy Gamma-ray Observatory experiment and comprises a silicon tracker, a cesium iodide calorimeter, and a scintillator anti-coincidence detector. All these will form the payload of a satellite. The detector will operate in the MeV range and provide a wider field of view than the Fermi-LAT detector. This detector is planned by NASA. e-Astrogram is a European Science Commission gamma-ray detector, based on similar instrumentation as AMEGO \cite{amego}. The e-Astrogram project aims to observe the frequency range of 0.3 MeV to 3 GeV. It is also aiming to be more sensitive at a particular frequency than previous instruments. These instruments will send data about the gamma-ray bursts and other sources which will give a clue on the existence of primordial black holes in the early universe.

\subsection{HESS}
The HESS is a gamma-ray observation experiment using an array of atmospheric imaging Cerenkov telescopes with energy in the TeV range. The telescopes are in Namibia. We report on the techniques of the HESS experiment in details here as an example, but it is one of several developments for PBH observations \cite{hess}. As the PBH which are smaller than $10^{17}$ g might have evaporated by now, one searches for gamma-ray burst signals. The PBHs are expected to have evaporated with an explosion of gamma rays, which have a high energy and last only for a few seconds. Using statistics and the methods of \cite{felco} Feldman and Cousins, one can estimate the ``rate of'' the PBH formation density $\dot{\rho}_{PBH}$, with 95\% and 99\% confidence levels. 
Further, we discuss this experiment's data analysis \cite{hess} in details to illustrate the methodology of the search of PBH. Let us say an unknown parameter $\mu$ is being assessed using the measurements of a variable $x$. Usually, one uses Bayesian statistics to estimate the ``belief'' in a system's parameter being $\mu_t$. This is given using the formula
\be
P(\mu_t|x_0)= {\cal L}(x_0|\mu)\frac{P(\mu_t)}{P(x_0)},
\ee
where ${\cal L}(x_0|\mu_t)$ is the ``likelihood'' of obtaining $x_0$ given $\mu_t$. However, it is assumed that there is prior knowledge of the probability $P(\mu_t)$ of finding $\mu_t$ independent of what $x_0$ is, which might not be the case. The probability $P(x_0)$ can be absorbed in the normalization of the conditional probability. In Bayesian methods, the belief in finding $\mu_t$ given the measured values of $x$ is expressed as a ``confidence''. This is mathematically
\be
\int^{\mu_2}_{\mu_1} P(\mu_t|x_0) d \mu_t=\alpha,
\ee
where $\alpha$ is the degree of confidence for $\mu_t$ to be in the confidence interval $[\mu_1,\mu_2]$. In \cite{felco}, a variation of this is given, for estimating the value of a parameter $\mu$ given the measurements of the variable $x$. 
If one takes the ratio of two likelihoods, then the ``prior knowledge'' required in Bayesian statistics is not there.
\be
R= \frac{{\cal L}(x|\mu)}{{\cal L}(x|\mu_{\rm best})},
\ee
where $\mu_{\rm best}$ is the value of the parameter which maximizes the conditional probability. This ratio determines the acceptance region in the $x$ variable, for a given value of $\mu$. A sum of the observation probabilities in decreasing order of $R$, until the required confidence limit is reached, provides a good estimate for the confidence intervals or upper limits for \mbox{a parameter. }

In the HESS observations, gamma rays were detected using the Cerenkov telescopes on earth. The number of photons detected could vary from one to infinity in a given time interval $\Delta t$. A time interval of $\Delta t=10$ s was taken for the purpose. We assumed that the detection of photon ``clusters'' of size $k$ followed a Poisson distribution 
\be
P(k, N)= e^{-N} \frac{N}{k!},
\ee
where $N(r,\alpha,\delta,\Delta t)$ is the number of $\gamma$ rays emitted from PBH from a distance $r$ in the angular interval in the sky specified by $\alpha,\delta$ in unit time $\Delta t$. Integrating this over all space, i.e., $r,\alpha,\delta$, and over all runs of the experiment, the number of significant clusters of photons detected were estimated to be
\be
n_{\rm sig}(k,\Delta t)= \dot{\rho}_{PBH} V_{\rm eff} (k, \Delta t),
\ee
where
\be
V_{\rm eff}(k, \Delta t)= \sum_i T_i \int d \Omega_i \int dr~ r^2~ P(k, N)= \sum_i T_i \Omega_i \frac{(r_0 \sqrt{N_0})^3}{2} \frac{\Gamma(k-3/2)}{\Gamma(k+1)},
\ee
where $N_0$ is the number of photons emitted from PBH at a distance of $r_0$. $T_i$ is the run's live time of the experiment, and $\Omega_i$ is the solid angle of the observations.
Based on the observed data, the statistical analysis using the techniques of Feldman and Cousins was implemented. The parameter being sought was $n_{\rm sig}$ given $n$ as the observed variable. Note that these photon clusters, which might be from evaporating PBH, were received along with the background photons, whose number was taken as $\bar{n}$, or off photons.
\be
R=\prod_{n}\frac{{\cal L}(n|\bar{n}+ n_{\rm sig})}{{\cal L}(n|\bar{n})}.
\ee

Here, the maximal value of the likelihood function was taken as that of the background $\bar{n}$. 

The $\chi^2$ estimate of the above can be found as \cite{felco}: 
\be
{\rm LNR} =-2\ln(R)= 2 \sum_{n} n_{\rm sig}+ n (\ln(n)- \ln(\bar{n}-n_{\rm sig})),
\ee
where $n$ is the number of observed photon signals in the on position of the telescopes and $\bar{n}$ is the number of mean observed signals in the off data. This is an estimate of the background photons, obtained by averaging over ``scrambled'' time intervals. In deriving the above, we used the Poisson distribution.

\begin{figure}
\centering
\includegraphics[scale=0.4]{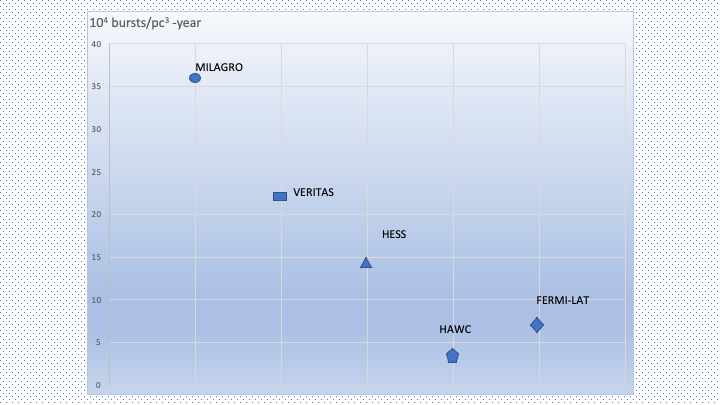}
\caption{The upper estimates of the number of final bursts at the 99\% confidence limit from some {experiments} 
\cite{hawc}.}
\label{fig:hawc}
\end{figure}

This ${\rm LNR}$ had a maximum of $0.006$ in the preliminary data for $\Delta t =10$ s and 6240 runs of four of the five telescopes \cite{hess}. This showed that there was not much of the PBH excess data. However, if one sets ${\rm LNR}=4,9$, one can obtain an upper-limit estimate for $\dot{\rho}_{PBH}$, with 95\% and 99\% confidence levels. The upper limit was found to be 
\bea
\dot{\rho}_{\rm PBH}& < &2.5 \times 10^{4} /pc^3 yr \  \ (95\%),\\
\dot{\rho}_{\rm PBH} &< & 5 \times 10^4 / pc^3 yr  \  \ (99\%),
\eea
These data points were further updated with other experiments such as VERITAS, MILAGRO, FERMI-LAT, and SWGO \cite{swgo}. {A comparative plot of the experimental predictions of evaporating PBH or final bursts at the 99\% confidence limit is given in Figure \ref{fig:hawc}. The data for this are quoted from \cite{hawc} (2021). For some recent updates in the field of constraints on PBH see \cite{rec}.}

For recent data on HESS, one can refer to the experiment's website \cite{hess1}.
\subsection{Neutrino Experiments}
The Hawking radiation from PBH releases neutrinos. The flux of these as a function of the PBH production and then a further analysis for ``secondary effects'' producing neutrinos were analyzed. The data from several experiments were taken and showed almost no or a very small estimation of the PBHs. Using a recent work \cite{neutrino}, we comment on the results. A neutrino spectrum rate was defined using the Hawking emission spectrum as in Equation (\ref{eqn:hawk}). Further, there can be secondary neutrino production due to the decay of hadrons produced initially:
\be
\frac{d^2 N_{\nu}}{d \omega_{\nu} d t}= \int_0^{\infty} dM \, \frac{d {\cal N}}{dM}\left( \frac{d^2 N_{\nu}}{d\omega_{\nu} dt}_{\rm prim} + \frac{ d^2 N_{\nu}}{d\omega_{\nu} dt}_{\rm sec}\right).
\ee
where the black hole's mass distribution could be taken as a Gaussian log-normal profile, 
\be
\frac{d {\cal N}}{ dM}= \frac{1}{\sqrt{2 \pi} \sigma M} \exp\left(-\frac{\ln^2(M/M_{\rm PBH})}{2 \sigma^2}\right),
\ee
or simply a delta function profile centred at $M=M_{\rm PBH}$. In the above, $M_{\rm PBH}$ is an average mass, and $\sigma$ is the standard deviation, as the mass of the black hole is allowed to vary.
A plot of the differential neutrino flux from extragalactic sources and the milky way can be calculated using publicly available software \cite{neutrino} and plotted. The differential flux of the neutrino plotted as a function of the energy $\omega_{\nu}$ varied between $10^2$ and $10^{-5}$, as the energy varied from 1 to 100 MeV for PBH of mass $10^{13}$ g. The evaporated PBH were taken as a fraction of the cosmic background which is $10^{-18}$ to obtain this result. 

The experimental bounds obtained from the Super-Kamiokande data showed that for PBH which were already evaporated, the abundance ratio was about $10^{-17}$ for $10^{13}$ g black holes and a confidence limit of 90$\%$. The question is of course what the above bounds imply for quantum gravity phenomenology? Whereas the PBH production cannot be ruled out completely, using the above estimation methods, it remains that the mechanism of PBH formation could be different, and the emission flux calculations greatly modified by intervening cosmic flows and quantum effects. In this aspect, one has to wait for future experiments such as JUNO, DARWIN, ARGO, and DUNE, and perhaps quantum cosmology predictions of the PBH formation from a more fundamental theory such as loop quantum gravity.  

It is obvious from the above discussions that the detection of bursts of photons and neutrinos on earth gives a very small window for the PBH to exist which would be evaporating now, i.e., those having masses $10^{5} {\rm g}$--$10^{14} {\rm g}$. However, as we know, there can still be the option that there are PBH which have not evaporated away but have formed remnants. These will still be candidate dark matter contributors. The fraction of PBH which contribute to dark matter and have not been evaporated yet is also estimated as $\sim 10^{-3}$ for masses of the order of $10^{16} {\rm g}$ as in \cite{Cedex}. There are other papers investigating this using various data sources such as microlensing, accretion disk luminosity, radio signals, anisotropies of the CMB, etc. We refer the reader to reviews in this field \cite{pbh}; there are also discussions of the PBH formation and evaporation using LQG corrected metrics, though in reduced phase-space formulations \cite{lqgred}. In our opinion, whereas the search is now much focused than earlier on what a gamma-ray burst or a neutrino flux from PBH may be, the research is still nascent.
\section{Event Horizon}
\label{event}
In the initial days of the discovery of the black hole metric solution to Einstein's equation, the existence of the horizon was one of the most bizarre predictions. The existence of trapped surfaces in general relativity was later firmly established with the Ray--Chowdhury equations and Hawking--Penrose singularity theorems. However, the debate continued on whether the event horizon existed, as it was unobservable. With the discovery of compact objects and the observation of X-rays from them, various models were tested for the existence of the event horizon. As the conclusions were model-dependent, the search continued, until the event horizon telescope project produced an ``assembled image'' of the photon sphere surrounding a black hole \cite{eht1, eht}. This confirmed some of the predictions about the behaviour of geodesics near a black hole's horizon, but did it confirm the presence of an event horizon? Perhaps not, but this is as ``good as it gets''. The snapshot of the photon sphere assimilated from eight infrared telescopes captured the electromagnetic waves circulating a compact object. The question we are asking in this article is: can we use the observations of geodesics around a black hole to measure semiclassical physics? In a work using semiclassical states in loop quantum gravity \cite{cohcorr}, it was shown that quantum fluctuations could cause instabilities in black holes, and these could produce tangible detectable effects for astrophysical black holes \cite{cohcorr}. The main results of the paper were the calculation of a nonpolynomial correction to the metric of the Schwarzschild black hole. The semiclassically corrected metric was shown to be of the following form
\bea
 ds^2  & = & -\left(1- \frac{r_g}{r} - \tilde{t} \ h_{tt} \right) dt ^2 + \tilde{t}\  h_{rt} \ dt dr  + \left\{\frac1{(1-r_g/r)} + \tilde{t} \ h_{rr} \right\} dr^2 + \nn \\ && +\left(r^2 + \tilde{t}  \ h_{\theta \theta} \right) d\theta^2 
  + \left(r^2 \sin^2\theta + \tilde{t} \ h_{\phi \phi} \right) d\phi^2. 
 \label{eqn:corrm}
\eea
where $r_g$ is the Schwarzschild radius, and the location of the horizon is at $r_g=2GM$, where $M$ is the mass of the black hole. $h_{tt}, h_{rt}, h_{rr},h_{\theta \theta},$ and $h_{\phi \phi}$ are the perturbations motivated from the corrections to the metric \cite{cohcorr}. The perturbations of the metric could be attributed to other quantum models of gravity, but we used the one motivated from \cite{cohcorr}, and a shift was generated, $h_{rt}$, breaking the ``static'' nature of the metric. The $\tt$ which appears in this coherent state was obtained using the length scales of the system and was thus a ratio of Planck's area to the area of the horizon $\tt= l_p^2/r_g^2$. Using this, we solved for the geodesics of the black hole. The geodesics were taken as circular orbits and the radial coordinate $r$ was solved as a function of the coordinate $\phi$.  These orbits described the trajectory of light rays which were incident on the black hole geometry from a distance, and the impact parameter measured the perpendicular distance of the light ray from the horizon. Using the invariant distance on the Schwarzschild geometry, one can write the equation of motion for the geodesic of a photon as a differential equation in the azimuth $\phi$, which was taken as the affine parameter along the geodesic.
{The deviations in geodesic computations for the rotating black hole from the nonrotating black holes were small \cite{eht3} but detectable. For rotating black holes, the cross section of the photon scattering might not be circular \cite{eht3}, but the difference was about 4\%. However, quantum corrections might be different, and one needs to formulate coherent states for rotating black holes separately. The effect of the presence of ``echoes'' might still be true. The results stated in this paper thus apply to nonrotating black holes strictly but pave the way for \mbox{realistic ones.}}

If we arrange the terms in a way they can be grouped into terms which are zeroth order in $\tt$ and then first order in $\tt$ (in the equatorial plane), one gets \cite{eht}:
{\small
\be
\frac{1}{r^4}\left(\frac{dr}{d\phi}\right)^2+ \frac1{r^2}\left(1-\frac{r_g}{r}\right)\left(1+ \tt \ \frac{h_{\phi \phi}}{r^2} - \tt \ h_{rr}\right) = \frac{1}{b^2}\left(1+ 2\ \tt \ \frac{h_{\phi \phi}}{r^2} -\tt \ h_{rr} +\tt \ \frac{h_{tt}}{1-r_g/r}\right).
\ee} 

\begin{figure}
    \centering
    \includegraphics[scale=0.8]{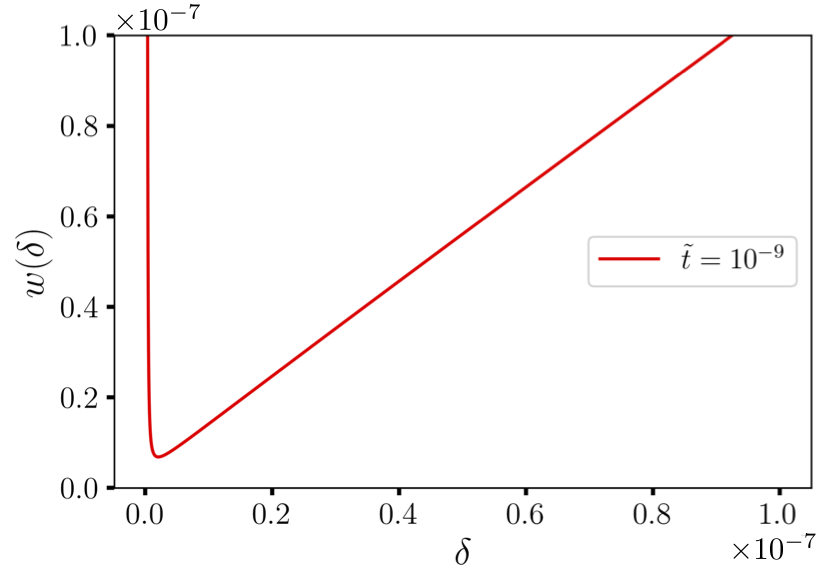}
    \caption{Plot of the semiclassically corrected photon geodesic impact parameter relation. The plot shows a bounce as the distance from the critical radius approaches the semiclassical length scale of $\tt\sim 10^{-8}$ units.}
    \label{fig:graph1}
\end{figure}

\begin{figure}
    \centering
    \includegraphics[scale=0.8]{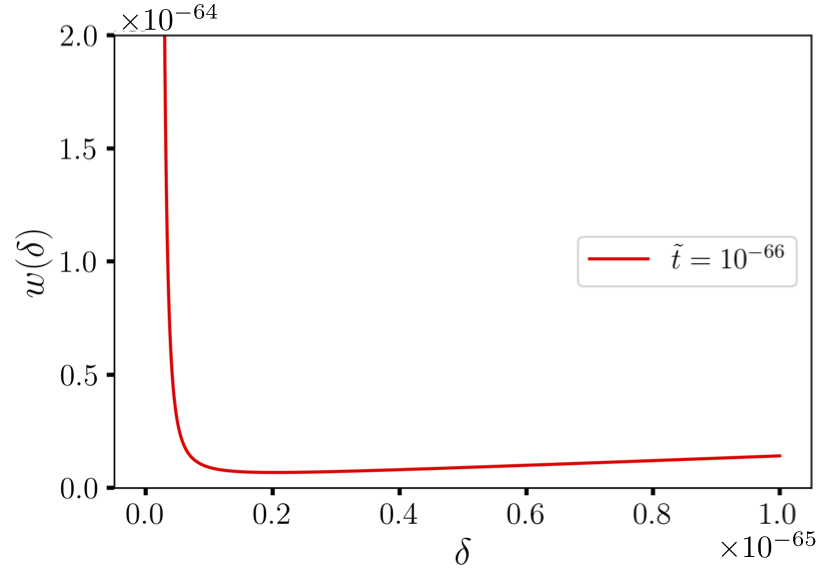}
    \caption{Plot of the semiclassically corrected photon geodesic impact parameter relation. The plot shows a bounce as the distance from the critical radius approaches the semiclassical length scale of $\tt\sim 10^{-66}$ units.}
    \label{fig:graph2}
\end{figure}

As one traces the trajectory through the entire path, the asymptotic angle of ``scattering'' from the black hole geometry emerges as a function of the impact parameter of the photon. 
The solution is obtained using a set of elliptic integrals and one finds
\be
\exp(-\phi_{\infty})= \delta^{1+0.0203~ \tt\ } \exp\left(+ \frac{0.47~ \tt^{1/2}\ }{(0.67 \delta +0.225 ~\tt)^{1/2}} +0.23~ \tt +1.712~ \frac{\tt}{\delta}\right),
\label{eqn:imp}
\ee
where $\delta= b-b_c$, and $\phi_{\infty}$ is the asymptotic angle the geodesic makes as it re-emerges to the asymptotic region. The difference of the photon geodesic impact parameter with the impact parameter of the critical orbit $b_c= 3\sqrt{3 } M$ is expected to be zero as the photon can orbit an infinite number of times round the horizon. 
One can see that in Equation (\ref{eqn:imp}), the $\tt \rightarrow 0$ reduces to a linear term in $\delta$. Most importantly, $\delta\rightarrow 0$ as $\phi_{\infty}= \mu + 2n \pi \rightarrow \infty$. $n$ counts the number of times the geodesic encircles the black hole, and this goes to infinity for the critical geodesic with the critical impact parameter. 
The photon circles the black hole an infinite number of times, when the critical impact parameter is reached.
If we take the semiclassical corrections, then the plot of $w(\delta)$ (the RHS of Equation (\ref{eqn:imp}) as a function of $\delta$ shows that the function does not reach zero but bounces off (see Figures \ref{fig:graph1} and \ref{fig:graph2}), and this we can associate with the presence of a quantization.   

This observation is commensurate with the work in fuzzballs and ECHOS \cite{ecos}. In these models, the horizon is replaced by a ``wall'' at a particular distance from the black hole. In our calculations with the LQG coherent states \cite{cohcorr}, we found the explicit location of the ``wall'' as a function of the semiclassical parameter $\tt$. We expect that our results can be eventually verified from observational data from astrophysical black holes \cite{ecos}.
\section{Conclusions}
As it happens, the search for quantum gravity in experiments is still nascent. However, we expect that in the early universe, the length scales were quantum, and therefore the search for relics of quantum gravity is ongoing. There are a number of papers in this Universe special issue in quantum gravity phenomenology which discuss cosmology and the effect of quantum cosmology in observational physics. In this review, the experiments we discussed only provided bounds on the mass of the graviton, the PBH production. We discussed the quantum effects which could be ``directly'' observable in recent experiments including in gravitational wave detectors and event horizon telescope images. 
We also reported on the numerous experiments which observe particles from distant celestial events on earth. We showed theoretical calculations and reported on bounds from experiments on Hawking emission from PBH. The experimental bounds did not violate any theoretical predictions. The observations provide directions for the experimental community to seek for more precise measurements. The plot of the electric and magnetic polarizations from the EHT \cite{eht2} and the launching of LISA \cite{lisa} are ongoing efforts in that direction. The study of fast radio bursts (FRB) provided an effort towards finding the quantum origins of astrophysical phenomena. The most promising experiments on earth for the quantum effects of gravity remain the GW detectors and the possibility that one would detect a ``graviton'' or its semiclassical version in the near future.

\vspace{6pt}

{\bf Acknowledgments}:This article is written for universe special issue in Quantum Gravity Phenomenology. AD would like to thank the universe editorial team, particularly Cici Xia for making this two volumes possible. AD is also thankful to co-editor Alfredo Iorio for collaboration.

\end{document}